\input harvmac
   
\Title{\vbox{\rightline{EFI-01-25}\rightline{hep-th/0010154 }}}
{\vbox{\centerline{Density of States and Tachyons} 
\centerline{in Open and Closed String Theory}}}
\medskip

\centerline{Vasilis Niarchos\footnote{$^*$}{vniarcho@theory.uchicago.edu}}
\medskip
\centerline{\sl Enrico Fermi Inst. and Dept. of Physics}
\centerline{\sl University of Chicago}
\centerline{\sl 5640 S. Ellis Ave., Chicago, IL 60637, USA}

\smallskip

\vglue .3cm
\bigskip 

\noindent

In this note we reexamine the possibility of constructing stable non-supersymmetric theories
that exhibit an exponential density of states. 
For weakly coupled closed strings there is a general theorem, according to which
stable theories with an exponential density of states
must exhibit an almost exact cancellation of spacetime bosons and fermions
(not necessarily level by level). We extend this result
to open strings by showing that if the above cancellation between
bosons and fermions does not occur, the open strings do not
decouple from a closed string tachyon even in the NCOS scaling
limit. We conclude with a brief comment on the proposed generalization
of the AdS/CFT correspondence to non-supersymmetric theories.

\Date{10/2000}

%
%

\def\frac#1#2{{#1\over#2}}

\def\ket#1{|#1\rangle}
\def\bra#1{\langle#1|}

\def\inbar{\,\vrule height1.5ex width.4pt depth0pt}
\def\IC{\relax\hbox{$\inbar\kern-.3em{\rm C}$}}
\def\IR{\relax{\rm I\kern-.18em R}}
\def\IP{\relax{\rm I\kern-.18em P}}

%
%

%
\catcode`\@=11
\def\slash#1{\mathord{\mathpalette\c@ncel{#1}}}
\overfullrule=0pt
\catcode`\@=12


%
\def\Tr{{ \rm Tr}}
%
\lref\ks{D.~Kutasov, N.~Seiberg ``Number of Degrees of Freedom,
Density of States and Tachyons in String Theory and CFT,''
Nucl.\ Phys.\ {\bf  B358}, 600 (1991)}
\lref\kut{D.~Kutasov ``Some properties of (Non) Critical Strings,''
hep-th/9110041}
\lref\KlebTsey{I.~R. Klebanov, A.~A. Tseytlin,
``D-Branes and Dual Gauge Theories in Type 0
Strings,'' hep-th/9811035}
\lref\gab{M.~R. Gaberdiel, ``Lectures on Non-BPS Dirichlet Branes,''
hep-th/0005029}
\lref\bergab{O. Bergman, M.~R. Gaberdiel, ``A non-supersymmetric
open string theory and S-duality,'' Nucl.\ Phys.\ {\bf B499}, 183
(1997), hep-th/9701137}
\lref\tseyzab{A.~A. Tseytlin, K. Zarembo, ``Effective potential in
non-supersymmetric SU(N)$\times$SU(N) gauge theory and interactions
of type 0 D3-Branes,'' hep-th/9902095}
\lref\sesutou{N. Seiberg, L. Susskind, N. Toumbas, ``Strings
in Background Electric Field, Space/Time Noncommutativity and 
A New Noncritical String Theory,'' hep-th/0005040}
\lref\gomamistr{R. Gopakumar, S. Minwalla, A. Strominger, ``S-Duality
and Noncommutative Gauge Theory,'' hep-th/0005048}
\lref\kleb{I.~R. Klebanov, ``Tachyon Stabilization in the AdS/CFT Correspondence,''
hep-th/9906220}
\lref\cole{S. Coleman, ``Aspects of Symmetry, Selected Erice Lectures of Sidney Coleman,''
Cambridge University Press, 1985}
\lref\wit{E. Witten, ``Baryons in the 1/N Expansion,''
Nucl.\ Phys.\ {\bf B160}, (1979)}
\lref\pol{J. Polchinski, ``String Theory, Vol. 1,'' Cambridge University Press}
\lref\gsw{M. B. Green, J. H. Schwarz, E. Witten, ``Superstring Theory, Vol. 2,'' 
Cambridge Monographs On Mathematical Physics}
\lref\banks{T. Banks, ``Cosmological Breaking of Supersymmetry or Little Lambda
Goes Back to the Future II,'' hep-th/0007146}
\lref\zamolo{A. B. Zamolodchikov, ``Conformal Symmetry and Multicritical Points
in Two-Dimensional Quantum Field Theory,'' Sov. J. Nucl. Phys. 44, 529 (1986)}
\lref\relev{J. Harvey, D. Kutasov, E. Martinec, ``On the relevance of tachyons,''
hep-th/0003101}
\lref\dienesone{K. R. Dienes,``Modular Invariance, Finitenes and
Misaligned Supersymmetry: New Constraints on the Numbers of Physical
String States,'' Nucl.\ Phys.\ {\bf B429}, (1994) 533,\ hep-th/9402006}
\lref\dienestwo{K. R. Dienes, M. Moshe, R. C. Myers, ``String Theory,
Misaligned Supersymmetry and the Supertrace Constraints,'' Phys.\ Rev.\
Lett.\ 74 (1995) 4767,\ hep-th/9503055}
\lref\wittwo{E. Witten, ``Instability of the Kaluza-Klein Vacuum,''
Nucl.\ Phys.\ {\bf B195}, (1982) 481}
\lref\klebants{I. R. Klebanov, A. A. Tseytlin, ``A Nonsupersymmetric Large N CFT from Type 
0 String Theory,'' JHEP {\bf 9903} (1999) 015, hep-th/9901101}
\lref\bankstwo{T. Banks, ``Matrix Theory,'' Nucl. Phys. Proc. Supp. 67 (1998)
180-224, hep-th/9710231}
\lref\sagnotti{M. Bianchi, A. Sagnotti, ``On the Systematics of Open
String Theories,'' Phys.\ Lett.\ {\bf B247} (1990) 517; A. Sagnotti,
``Some Properties of Open-String Theories,'' hep-th/9509080;
``Surprises in Open-String Perturbation Theory,'' Nucl.\ Phys.\ Proc.\
Supp.\ {\bf B56} (1997) 332, hep-th/9702093}


\noindent{\bf 1. INTRODUCTION} 
\medskip

In field theory one can pose the following question.
Is it possible to find stable, non-supersymmetric, weakly interacting
theories with an exponential density of states?
This is an interesting question for a number of reasons.
For example, QCD appears to be an example of this type of theories.
In the limit of a large number of colors it contains an infinite
number of weakly interacting mesons \refs{\cole,\wit}. 
These states are expected to lie
on Regge trajectories and their density to be exponential. 
A different reason is the more general question of 
supersymmetry breaking in string theory and in particular
the problem of the cosmological constant.

String theories generically possess a Hagedorn spectrum of states
and therefore offer a suitable arena to discuss 
the above questions. The situation for closed string theories
has been analyzed in \refs{\ks,\kut} 
(see also \refs{\dienesone,\dienestwo} for a related discussion)
and has been 
found that in the context of a generic 
weakly coupled closed string theory the existence of tachyon
instabilities and the density of states are related.
More precisely, 
whenever a string theory is tachyon free, 
the number of bosons almost cancels the number of fermions.
Following \refs{\ks,\kut} we refer to this almost exact cancellation
by using the term {\it asymptotic supersymmetry}. 
A special case of this phenomenon occurs for supersymmetric theories,
because supersymmetry guarantees the exact matching of the number of
bosons and fermions in the physical spectrum. In general, however, 
spacetime supersymmetry is not necessary for a stable theory 
with an exponential density of states and this
matching may or may not be strictly exact. The necessary property is
asymptotic supersymmetry. We review the relevant arguments in section 2.     

Along these lines,
it is natural to ask whether a similar situation appears in open
string theories as well.
Naively, this does not seem to be the case,
because there are known examples of string theories 
with open string spectra that include an exponential density
of states without
the above boson-fermion cancellation and without an open string 
tachyon. 
As we are going to see, however, an open string spectrum with an
exponential density of states and no asymptotic supersymmetry cannot be
IR stable in the weak coupling regime, because it cannot be decoupled
from a {\it closed} string tachyon even in the NCOS (noncommutative
open string) scaling limit. Hence, asymptotic 
supersymmetry is needed in open string theories as a condition for
stability of the bulk and not as a condition
for stability of the branes.

As an example of this general phenomenon, we discuss D-branes in the type
0 theories and comment on the proposed AdS/CFT correspondence in this class of 
theories. 

\vskip .5cm
\noindent{\bf 2. CLOSED STRING THEORY} 
\medskip
We mentioned above that in weakly coupled closed string theory, there is a
general
theorem relating the existence of tachyon instabilities with
the density of states. Following \refs{\ks,\kut} let us briefly
review this result. Consider a generic closed string vacuum. The
torus partition function takes the general form
\eqn\asd{
Z_T(\tau)=\Tr _{\cal{H}_{\rm c}}
q^{L_0-\frac{c}{24}}\overline q^{\overline L_0-\frac{c}{24}}
}
where $q=e^{2\pi i \tau}$ and $\tau=\tau_1+i\tau_2$ is the worldsheet
modulus. In terms of this partition function the one-loop
free energy $\Omega$ equals
\eqn\dgf{
\Omega=\int_{\cal{F}} \frac{d^2\tau}{\tau_2^2} Z(\tau)
}
where $Z(\tau) \equiv \tau_2 Z_T(\tau)$ and $\cal{F}$ is the
fundamental domain of the moduli space of the torus.
In field theory the corresponding amplitude has a UV divergence
and needs to be regularized, but in closed string theory this divergence
is cut off automatically by restricting the integral to the fundamental 
domain. Thus, the only possible
divergence of $\Omega$ is an IR one (coming from $\tau_2 \rightarrow \infty$)
and is associated with the presence of tachyons in the physical spectrum of
the theory.

Now define the following function
\eqn\fhg{
G(\tau_2) \equiv \int_{-\frac{1}{2}}^{\frac{1}{2}} d\tau_1 Z(\tau)
}
The integration over $\tau_1$ correctly implements the level
matching condition $L_0=\overline L_0$ for the left and right movers.
In terms of the Coleman-Weinberg formula, which gives $\Omega$
as
\eqn\jki{
\Omega=\Tr (-1)^F \log(p^2+m^2)=
\int_0^{\infty} \frac{ds}{s} \sum_n (-1)^{F_n} \int
d^D p e^{-s(p^2+m_n^2)}
}
the $G$ function reads 
\eqn\thb{
G(\tau_2)=\tau_2\sum_n (-1)^{F_n} \int d^D p \ e^{-2\pi \tau_2 
(p^2+m_n^2)}
}
From this formula we see in particular that spacetime fermions contribute
with a minus sign and therefore it is natural to write $G$ as
\eqn\kicnh{
G=G_B-G_F
}
In the $\tau_2 \rightarrow 0$ limit both $G_B$ and $G_F$ take
the general form
\eqn\wervgj{
G_{B,F} \simeq \tau_2^x e^{y/\tau_2}
} 
and can be thought of as regularized versions of the number of states
of the system. In particular, $G$ can be thought of as a measure of the difference
between the number of bosonic and fermionic degrees of freedom in the
physical spectrum of the theory.

The theorem of references \refs{\ks,\kut}
postulates that under very mild
assumptions the following relation holds
\eqn\kweu{
\Omega=\int_{\cal{F}} \frac{d^2 \tau}{\tau_2^2} Z(\tau)=
\frac{\pi}{3} \lim_{\tau_2 \rightarrow 0} G(\tau_2)
}
In any infrared stable theory
the l.h.s. of this equation is finite. It follows then that 
\eqn\evy{
\lim_{\tau_2 \rightarrow 0} G(\tau_2)=\lim_{\tau_2 
\rightarrow 0} (G_B(\tau_2)-G_F(\tau_2))=\rm{const}
}
Hence, in any tachyon-free theory bosons and fermions in the
physical spectrum almost cancel.
This cancellation takes place in the asymptotic high energy limit $\tau_2
\rightarrow 0$ and 
may not be exact or level by level, but
the total difference of states up to some high energy level must
asymptotically approach the low residual density
of states counted by $G$. In fact, as explained in \refs{\ks,\kut},
the $G$ function converging to a constant in the $\tau_2 \rightarrow 0$ limit
corresponds to a 2-dimensional field theoretic density of states.  
This phenomenon of asymptotic cancellation between the number of bosons 
and fermions up to a 2-dimensional
density of states is what we refer to as asymptotic supersymmetry.
The conclusion is that 
any stable weakly coupled closed string theory, if not supersymmetric, has
to be at least asymptotically supersymmetric.

From this point of view closed string tachyon condensation proceeds in the 
following way. Suppose we start with 
an unstable weakly coupled closed string theory. Then, the spectrum of
this theory does not have asymptotic supersymmetry and includes for
example an exponential density of purely bosonic degrees of freedom.
After tachyon condensation there are two possibilities. Either the
new theory is strongly coupled, in which case we cannot say
anything from the point of view of our analysis, or the new theory is
still weakly coupled. In that case, the new vacuum must be
asymptotically supersymmetric. That means that either an exponential 
density of fermions has dynamically appeared to match asymptotically the
density of bosons, or more drastically spacetime collapses down to two
dimensions by some violent mechanism (probably of the sort encountered in
\refs{\wittwo}).  This qualitative picture also seems to be in agreement
with the picture one has about closed string tachyon condensation from the
point of view of the worldsheet. From that point of view the closed string
tachyon corresponds to a relevant perturbation on the worldsheet and under
the renormalization group the worldsheet theory flows from the UV to the
IR. Because of the c-theorem, the c-function will have a smaller value in
the IR and this suggests either that fermionic degrees of freedom have
appeared in the new stable vacuum, or that the spacetime dimension has
been drastically reduced.

\vskip .5cm

\noindent{\bf 3. OPEN STRING THEORY} 
\medskip
For closed string theory modular invariance was critical to the proof
of the above theorem. For open strings analogous results
can be obtained by using worldsheet duality.
Open string theory generally possesses different sectors, but for the moment
let us restrict our discussion to the case of an
$ab$ sector, corresponding to boundary conditions
$\ket{a},\ket{b}$. For this sector we consider the 
annulus partition sum
\eqn\wufg{
\Omega_{ab}=\int_0^{\infty} \frac{dt}{t} \Tr_{{\cal H}_{ab}}
q^{L_0-\frac{c}{24}}
=\int_0^{\infty} \frac{dt}{t} A_{ab}(t)
}
where $q$ is now redefined to be $q=e^{-2\pi t}$ and 
$A_{ab}(t) \equiv \Tr_{{\cal H}_{ab}} $$q^{L_0-\frac{c}{24}}$. Again, the
Coleman-Weinberg formula \jki\ gives
\eqn\uvb{
A_{ab}(t)=\sum_{n \in ab} (-1)^F \int d^Dp \ e^{-2\pi t(p^2+m_n^2)}
}
where by $\sum_{n \in ab}$ we signify a sum over the open string 
states of the $ab$ sector.
In parallel to the closed string case, the $t \rightarrow 0$
limit of $A_{ab}=A_{B,ab}-A_{F,ab}$ counts the asymptotic difference
between the number of bosonic and fermionic degrees of freedom 
within the $ab$ sector and in general both $A_{B,ab}$ and $A_{F,ab}$
take (in this limit) the form \wervgj\
\eqn\oiuffg{
A_{B,F}(t) \simeq t^x e^{y/t}
}

As we know \refs{\pol,\gsw}, we may 
use worldsheet duality to
rewrite $A_{ab}$ in terms of the
transverse closed string channel variable $\tilde q=e^{-\frac{\pi}{t}}$
\eqn\qwerv{
A_{ab}(t)=2t\bra{b}\tilde q^{\frac{1}{2}(L_0+\overline L_0-\frac{c}{12})}
\ket{a}
}
and reinterpret the above UV divergences (from the $t \rightarrow 0$ limit) 
as IR effects associated to the
propagation of closed string modes. 
An exponential divergence signals the existence of a closed string tachyon 
with a nonzero coupling to the open strings. 
This tachyon has a negative mass squared $-\frac{2y}{\alpha'}$
in direct relation to the power of the exponential divergence in \oiuffg.
Similarly, the next dominant term (or the first dominant term
if there are no tachyons in the bulk) is a negative power that
corresponds to the propagation of massless closed string modes. 
This power 
\eqn\lekfig{
A_{ab}(t) \simeq t^x
}
can be determined in the following way.
For momentum $k^{\mu}$ flowing between the cylinder boundaries,
the annulus amplitude \wufg\ also includes
a factor $e^{-\alpha'k^2s/2}$ in the large $s=1/t$ limit. 
Of course, without any external open
string insertions this momentum is zero, but we can insert it as a regulator
into the annulus partition sum
\eqn\wgtnb{
\Omega_{ab} \simeq \int_0^{\infty} \frac{dt}{t}t^x e^{-\frac{
\omega}{t}}
}
and afterwards take the limit
$\omega=\frac{1}{2}\alpha' k^2 \rightarrow 0^+$.
After a change of variables 
this amplitude becomes
\eqn\sdfkww{
\Omega_{ab} \simeq \int^{\infty}_0 \frac{ds}{s} s^{-x} e^{-\omega s}=
\omega^x\int^{\infty}_0 dz z^{-x-1}e^{-z} = (\frac{1}{2}\alpha'k^2)
^x \Gamma(-x) 
}
From the field theory point of view 
this result should be proportional to the momentum
pole $\frac{1}{k^2} \bigg | _{k^{\mu}=0}$.  
Clearly, this can only happen when $x=-1$, in which case,
as we take the limit $\Lambda^{-2} \simeq t \rightarrow 0$, $A_{ab}(\Lambda) \simeq 
\Lambda^2$ and therefore 
corresponds to a 2-dimensional field theoretic density of states.

In summary, we conclude that in any given open string
sector a net exponential density of states yields a nonzero coupling
of the open strings to a closed string tachyon. On the other hand, the
absence of this coupling to a closed string tachyon implies a cancellation
between the bosonic and fermionic degrees of freedom up to a
2-dimensional density of states. This result is analogous to
the one that has been obtained in the closed string case, but 
there are also some major differences. In the closed string case
the stability of the theory was synonymous to asymptotic supersymmetry,
but in the open string case instabilities can be due to either
an open or a closed string tachyon. It is the absence of a coupling to 
a closed string tachyon that is equivalent with asymptotic supersymmetry. Open string 
tachyons, instead, signal
the existence of lower energy configurations for the
D-branes in question. This observation seems to be implying something deep about the
importance of asymptotic supersymmetry in weakly coupled string theories. 
Its presence, even for those degrees of freedom that live on the branes,  
appears to be intimately connected to the nature of spacetime. This phenomenon has
also been observed in matrix theory, where asymptotic supersymmetry seems to be a
crucial requirement for locality and cluster decomposition \refs{\bankstwo}.

The same arguments can be applied beyond the spectrum of a particular
open string sector $ab$. To implement our conclusions for a collection
of open sectors we would have to consider the annulus partition sum
\eqn\weik{
\Omega_A=\sum_{ab} \Omega_{ab}=\int \frac{dt}{t} A(t)
}
where $A(t) \equiv \sum_{ab}A_{ab}(t)$ and the summation is performed
over all the sectors we are considering. Apart from this minor modification
the preceding discussion goes through unchanged.

\vskip 0.4cm
\noindent{\it{The noncommutative open string case}} \rm
\medskip   
It has been argued in a recent series of papers \refs{\sesutou,\gomamistr},
that there are consistent open string theories without
a coupling to the closed sector.
If a full decoupling of this sort were always possible, tachyon free open
string theories with an exponential density of states and no
asymptotic supersymmetry would be examples of the stable
theories discussed in the introduction. 
But these are precisely the kind of string theories we excluded by
the above considerations.
The resolution to this problem is rather straightforward.
A careful examination 
of the decoupling argument in the NCOS shows that
the closed string tachyon and a 2-dimensional portion of the massless closed
string spectrum 
do not in general decouple from the open sector.

The idea behind the NCOS construction is to start with an ordinary
string theory of open plus closed strings and take a particular
scaling limit by turning on a near critical electric field. The resulting
open string theory has the same spectrum and the same annulus partition 
sum as before but it is 
described by a new effective string length given by the fixed 
value $\alpha'_{\rm{eff}}=\frac{\theta}{2\pi}$, $\theta$ being
the noncommutativity parameter, a new open string coupling
$G_o$ and a fixed open string metric $G_{MN}$, which can
be set to $\eta_{MN}$. 
At the same time, the closed
string coupling $g_s$ is scaled to infinity, whereas the closed
string metric is given by the form $g_c=\bigg ( \matrix{g&0\cr
                                                     0&-g\cr} \bigg)
\otimes 1_{(D-2) \times (D-2)}$ and $g$ is also sent to infinity.
For simplicity, we have assumed that the branes have only one
common direction and that the near critical electric field is turned
on in this direction, which we set as the direction 1.
In this context,                                                       
consider the process of an open string turning into  
a closed string. The open string dispersion relation reads
\eqn\zxvn{
(p^0)^2=(p^1)^2+m_o^2
}
and the closed string dispersion relation
\eqn\qwcbmg{
\frac{1}{g}(p^0)^2=\frac{1}{g} (p^1)^2+(\vec{p})^2+m_c^2
}
For $g \rightarrow \infty$ we get
\eqn\dhgj{
0=(\vec{p})^2+m_c^2
}

If $m_c^2>0$, the above relation cannot be satisfied and the respective closed
string modes decouple, as in the original argument. 
This is not the
case, however, for $m_c^2 \leq 0$. In particular, the tachyon has $m_c^2<0$
and therefore does not decouple by the above kinematic argument.
Hence, even in the NCOS scaling limit, an open string theory
with an exponential density of states and no asymptotic supersymmetry
cannot avoid coupling to a closed string tachyon and therefore
must be IR unstable.
Furthermore, for $m_c^2=0$, the massless modes too do not completely 
decouple. 
Kinematically, the modes that remain coupled to the open sector
are confined in the direction of the
electric field and become effectively 2-dimensional degrees of 
freedom. Of course, gravity decouples completely, since
in our limit $\alpha' \rightarrow 0$ and gravity
always decouples at low energy.



\vskip 0.4cm

\noindent{\it{An example}} \rm
\medskip
As a further illustration, we would like to present 
an example. For our purposes, a convenient setup is provided by
the type $0$ theories. The D-brane spectrum of these theories
has been analyzed and has been found to consist of two
types of mutually consistent branes plus their anti-branes.
\foot{Open string descendants of the 0B type were constructed
by orientifold projection in \sagnotti .} 
Following \refs{\gab,\bergab} the first type, which will be called electric,
is given by the boundary state
\eqn\wmxufh{
\ket{Dp;+,+}=\ket{Dp,+}_{NSNS} +\ket{Dp,+}_{RR}
}
whereas the second type, which will be called magnetic, is given by 
\eqn\egvjr{
\ket{Dp;-,-}=\ket{Dp,-}_{NSNS}+\ket{Dp,-}_{RR}
}
The +,$-$ signs in this notation refer to the choice of different
spin structures one can use to define the gluing conditions
satisfied by these boundary states. A detailed construction
can be found in \refs{\gab}.

The open string spectrum of these branes is the following.
Strings stretching between branes of the same
type belong to the $NS^+$ sector, whereas strings stretching between
branes of different type belong to the $R^+$ sector.
For this theory a generic configuration of parallel D-branes exhibits a tachyon-free 
open string spectrum with an exponential density of 
states and no asymptotic supersymmetry.
As one might expect, these branes also have 
a nontrivial coupling to a closed string tachyon 
(see references \refs{\KlebTsey,\kleb}) and this
coupling will also persist in the NCOS scaling limit by virtue
of the above discussion. 

In addition, it has been noted in \refs{\KlebTsey,\kleb,\tseyzab}
that the above coupling to a bulk tachyon vanishes for a dyonic pair
consisting of an electric plus a magnetic brane. 
From our point of view, this is clearly expected. On a dyonic
pair of branes the open string spectrum is 
asymptotically supersymmetric and the closed string tachyon decouples.
In fact, it is worth noticing that in this case 
the number of bosons exactly matches the
number of fermions level by level, despite the absence 
of spacetime supersymmetry \klebants.
In \refs{\KlebTsey,\kleb} it was argued that
for weak t'Hooft coupling such a D-brane system exhibits
a closed string tachyon stabilization and that
AdS/CFT correspondence in this non-supersymmetric setting makes sense. In
that argument,
however, tachyon condensation is a tree-level effect at weak coupling and
the bulk spectrum continues to have no asymptotic
supersymmetry. Thus, it is not clear 
how such a mechanism could be reconciled
with the analysis of section 2.
This difficulty with the tachyon is even greater
for a stack of $N$ branes of the same type. 
The closed string tachyon in the bulk cannot be stabilized
in the weak coupling regime because it
must remain coupled to the $U(N)$ gauge theory on
the branes. Therefore, 
in such a generalization of the AdS/CFT
correspondence to non-supersymmetric theories one cannot neglect 
the closed string tachyon on the ``CFT'' side. 

\vskip .5cm
\noindent{\bf 4. CONCLUSIONS}
\medskip

In this note we posed the following question. Is it possible to 
find stable, non-supersymmetric, weakly interacting theories
with an exponential density of states? 
Probing this question in the context of perturbative string theory, we
commented on the correlation between tachyon
instabilities and exponential densities of states without
asymptotic supersymmetry. We saw that, as in the case of
closed string theories, the density of states of an open string spectrum
dictates the coupling of the open strings to closed string
tachyons. It was stressed that this coupling is not an
accidental fact that can be removed by decoupling the closed
sector. Even in the NCOS scaling limit this coupling
remained as needed by the above discussion on the density
of the open string states.

A natural question to ask is the
following. What is the meaning of these conclusions for large $N$ 
QCD? Clearly, there are three possibilities. One possibility is that
large $N$ QCD presents an example of a stable, weakly coupled
theory with an exponential density of states. In that case,
however, the discussion of this note shows that it
cannot be described by a weakly coupled string theory of the usual kind and
one must find a different kind of string theory to describe it.
Another alternative is that large $N$ QCD
does not present an example of the above type of theories,
i.e. either it does not
have an exponential density of states or it is not weakly coupled.

Finally, in recent years some evidence appeared for a deep connection
between asymptotic supersymmetry and the emergence of large
spacetimes with approximately local physics in string theory
(see e.g. \bankstwo\ or for a recent discussion \banks\ ). The analysis in
this note is compatible with these ideas. Indeed, consider
a vacuum of string theory which contains D-branes. If the
spectrum of states that live on the D-branes is not asymptotically
supersymmetric, the vacuum must suffer from a closed string
tachyon instability, and spacetime (even very far from the
brane) will collapse in the sense that was described at the end of section 2. This 
seemingly counter-intuitive
relation between large breaking of supersymmetry on a brane,
and instability of the bulk of spacetime might be a further
sign of the above connection.


\vskip 1cm
\noindent{\bf Acknowledgments}

\

\noindent{I would like to thank D. Kutasov for many useful discussions
and especially for his encouragement and support. Also, I would like to 
thank N. Prezas for several remarks on the second version of this paper.
This work was supported by DOE grant DE-FG02-90ER40560.}

\listrefs
\end